\documentclass{PoS}
\usepackage{amsmath}
\usepackage{amsfonts}
\newcommand{\mobius}{M\"obius~}
\newcommand{\ua}{${\rm U}(1)_\text{A}$ }
\newcommand{\Tr}{{\rm Tr}}

\usepackage[utf8]{inputenc}

\title{\begin{picture}(0,0)(0,0)%
    \put(350,75){\makebox(0,0)[l]{\textnormal{\normalsize KEK-CP-330}}}%
    \put(350,60){\makebox(0,0)[l]{\textnormal{\normalsize OU-HET-878}}}%
  \end{picture}%
  On the axial $U(1)$ symmetry at finite temperature}

\ShortTitle{On the axial $U(1)$ symmetry at finite temperature}

\author{JLQCD Collaboration: \speaker{Guido Cossu}$^a$\thanks{E-mail: cossu@post.kek.jp}, Hidenori Fukaya$^c$, Shoji Hashimoto$^{a, b}$, Jun-ichi Noaki$^a$, Akio Tomiya$^d$\\
\\
\llap{$^a$}Theory Center, IPNS, High Energy Accelerator Research Organization
  (KEK), Tsukuba 305-0801, Japan\\
\llap{$^b$}School of High Energy Accelerator Science, The Graduate University for Advanced Studies (Sokendai), Tsukuba 305-0801, Japan\\
\llap{$^c$}Department of Physics, Osaka University, Toyonaka 560-0043, Japan\\
\llap{$^d$}Key Laboratory of Quark \& Lepton Physics (MOE) and Institute of Particle Physics, Central China Normal University, Wuhan 430079, China}

\abstract{We study the \ua anomaly in two-flavor lattice QCD at finite temperature with the \mobius domain-wall Dirac operator.
We generate gauge configurations in the temperature range $(0.9, 1.2) T_c$ on different physical volumes, $L=$ 2--4 fm, and lattice spacings.
We measure the difference of the susceptibilities of the flavor non-singlet scalar ($\chi_\delta$) and pseudoscalar ($\chi_\pi$) mesons. They are related by an axial $U(1)$ transformation and the difference vanishes if the axial $U(1)$ symmetry is respected. 
We identify the source of axial $U(1)$ symmetry breaking  at finite temperature in the lowest eigenmodes, for the observable $\chi_\pi - \chi_\delta$.
We then reweight the \mobius domain-wall fermion partition function to that of the overlap-Dirac operator to fully recover chiral symmetry. 
Our data show a significant discrepancy in the results coming from the \mobius domain-wall valence quarks, the overlap valence quarks on our DWF configurations and the reweighted ones that have full chiral symmetry. 
After recovering full chiral symmetry we conclude that the difference $\chi_\pi - \chi_\delta$ shows a suppression in the chiral limit
that is compatible with an effective restoration of \ua at $T \gtrsim T_c$ in the scalar meson channels.  }

\FullConference{The 33rd International Symposium on Lattice Field Theory\\
		14 -18 July 2015\\
		Kobe International Conference Center, Kobe, Japan*}

\begin{document}

\section{Introduction}

The question whether the \ua symmetry is effectively restored above the chiral phase transition is still open. In the well-known pattern of symmetry breaking in $N_f$ flavor QCD at low temperature 
\begin{equation}
SU(N_f)_L \otimes SU(N_f)_R \otimes U(1)_V \otimes U(1)_A \rightarrow U(1)_V \otimes SU(N_f)_V ,
\end{equation}
the \ua symmetry is peculiar since it is violated by the quantum anomaly. It comes from the presence of topological fluctuations that generate an anomalous contribution to the divergence of flavor-singlet axial-vector current \cite{'tHooft:1976up}.

The answer to this question may have an impact also from a  phenomenological viewpoint: the order and the universality class of the phase transition depend on whether the axial symmetry is restored or not \cite{Pisarski:1983ms,Pelissetto:2013hqa}.
Models like the instanton gas \cite{Gross:1980br} predict a suppression of the instanton density and thus an effective restoration of the \ua symmetry at very high temperatures $T \gg T_c$\footnote{$T_c$ is the temperature of the chiral phase transition, namely the location of the peak of the susceptibility of the chiral condensate.}, in the domain of their applicability. Only recently the lattice QCD studies on this subject have been (re)started at around the phase transition using several formulations of the fermion action and focusing on  different observables \cite{Cossu:2013uua,Cossu:2014aua,Tomiya:2014mma,Buchoff:2013nra,Brandt:2013mba,Dick:2015twa,Chiu:2013wwa}.

In the previous JLQCD work we studied the problem using the overlap fermion formulation \cite{Cossu:2013uua}. This guarantees exact chiral symmetry of the lattice action in the chiral limit. The Dirac spectrum and the meson correlator measurements both indicate a restoration of the \ua symmetry in QCD with two degenerate flavors. A gap in the spectrum opens at temperatures above $T_c$ when the quark mass is decreased toward the chiral limit. At the same time, the disconnected diagrams vanish, leading to a degeneracy of the correlators of the lightest mesons, which is a signal of the restoring symmetry. The problem was also studied theoretically in \cite{Aoki:2012yj} showing that with two degenerate flavors the spectral density of the Dirac operator behaves like $\rho(\lambda) \sim c\lambda^3$ in the high temperature phase. It implies that the \ua anomaly is invisible in the meson susceptibilities. This result is compatible with our lattice simulations.  

The most important source of systematic errors in the previous project was the need to fix the global topology $Q$. In order to avoid this limitation we started a new series of simulations using the M\"{o}bius domain-wall fermion formulation \cite{Kaneko:2013jla} with the code platform IroIro++ \cite{Cossu:2013ola}. Compared to the standard domain-wall formulation we have the advantage of having smaller residual mass, i.e. better chiral symmetry. As we are showing in these proceedings, a precise chiral symmetry is quite relevant for the study of the \ua problem and even M\"{o}bius fermions would not be sufficient. Another important issue is the mass dependence: we only observe the restoration when approaching the chiral limit. The current results are in accordance with the outcome of the previous overlap project.

In the following sections we present the methodology of our analysis and discuss the results that will be discussed in detail in an upcoming paper in preparation.

\section{Analysis}

We study $N_f=2$ QCD with the tree-level Symanzik improved gauge action and smeared M\"{o}bius domain-wall fermions. The details of this fermonic action are reported in \cite{Kaneko:2013jla} and are the same as our zero temperature simulations \cite{Noaki:2014ura}. Simulation points cover a region of temperatures between 150 and 250 MeV with up to three different masses for the points just above the phase transition. The measured residual mass above the phase transition is $O(1)$ MeV for the coarser runs, $N_t=8$, and less than half a MeV for the finer runs, $N_t=12$~\cite{CossuArticle}.

We measure two main observables related to the axial $U(1)$ symmetry: the eigenvalue spectrum $\rho(\lambda, m)$ of the hermitian Dirac operator ($H \equiv \gamma_5 D$) and the \ua susceptibility $\Delta$ defined as a difference of the susceptibilities of $\pi$ and $\delta$ channels

\begin{equation}
\Delta = \chi_\pi - \chi_\delta = \int {\rm d}^4x \langle \pi^a(x) \pi^a(0) - \delta^a(x)\delta^a(0) \rangle.
\label{eq:Delta}
\end{equation}
It vanishes when the \ua symmetry is fully restored in the vacuum. This quantity has a simple representation in terms of the Dirac operator eigenvalue spectrum:

\begin{equation}
\Delta = \lim_{m \rightarrow 0}\lim_{V\rightarrow \infty} \int \frac{2m^2 \rho(\lambda,m)}{(\lambda^2+m^2)^2} {\rm d}\lambda = \lim_{m \rightarrow 0}\lim_{V\rightarrow \infty} \Bigl(\frac{2N_0}{Vm^2} + \sum_{\lambda_i \neq 0} \frac{2m^2(1-
\lambda_i^2)^2}{V(\lambda_i^2 + m^2)^2(1-m^2)^2} \Bigr ),
\label{eq:DeltaSpectr}
\end{equation}
where the limits must be taken in that order and $N_0 = |Q|$, the number of zero eigenvalues of $D$. The second equation comes from an expansion in the eigenvalues $\lambda_i$ of the discretized overlap operator. Notice that the term depending on the zero modes is expected to vanish in the thermodynamical limit since a constant topological susceptibility $\chi_t=\langle Q^2 \rangle/V$ implies that $N_0 / \sqrt{V}$ is constant, thus  $N_0 / V \rightarrow 0$ when $V \rightarrow \infty$\footnote{Empirically, on a finite lattice the total number of zero modes is always equal to the number of left handed or right handed modes}.

Another useful representation of the susceptibilities is obtained using traces of the Dirac operator propagators. In terms of the massive 4D quark propagator ${D}_m^{-1}$, they are written as
\begin{equation}
  \chi_\pi = \frac{1}{V} \Tr [(\gamma_5 D_m)^{-2}], 
  \quad 
  \chi_\delta = - \frac{1}{V} \Tr[ (D_m)^{-2}],
  \label{eq:SuscTraces}
\end{equation}
after averaging over the source point. 

We concentrate on $\Delta$ and the discussion on the spectrum of the Dirac operator is given in another paper of these proceedings \cite{NewAkio}.

\section{Results}

In this section we discuss the measurements of $\Delta$, eq.~(\ref{eq:Delta}), using domain-wall fermions. We first show that $\Delta$ can be seriously affected by lattice artifacts coming from the violation of the Ginsparg-Wilson relation, $\{\gamma_5, D\} = 2 aD\gamma_5 D$, on coarse lattices. In order to eliminate this discretization effect we discuss a procedure to reweight the partition function of the domain-wall fermions to that of the overlap fermions that guarantee exact chiral symmetry. We then discuss the results of this procedure on the $\Delta$ measurement. 

The measurement of $\Delta$ is quite delicate in many aspects and the details of the method could affect the final result. We observe that a simple integration of the correlator from a local source is highly sensitive to the position of the source, and this needs high statistics. This is explained by the spatial location of the zero and lowest-lying modes of $H$. A source hitting one of these modes would overestimate the final result and viceversa if far away. A stochastic estimate of eq.~(\ref{eq:SuscTraces}) using a $Z_2$ noise source all over the volume is more reliable in this respect. 

\subsection{Ginsparg-Wilson relation violation}

Taking into account a possible operator $\mathcal V$ that represents the violation of chiral symmetry, the Ginsparg-Wilson relation is rewritten as
\begin{equation}
  \{\hat \gamma_5, H_0 \} = \mathcal V ,
  \label{eq:GWH}
\end{equation}
with $\hat\gamma_5\equiv\gamma_5-H_0$, and $H_0 \equiv \gamma_5 D^{4d}_\text{DW}(0)$ the massless hermitian Dirac operator. The spectral decomposition of the susceptibilities can then be written including the effect of the matrix elements $ \mathcal V_{nk}=\langle\psi_n| \mathcal V|\psi_k\rangle$, where $|\psi_n\rangle$ satisfies $H_m |\psi_n\rangle = \lambda_n |\psi_n\rangle$, an eigenmode of the massive operator $H_m$. The details of this decomposition are discussed in \cite{Cossu:2015kfa}. We report the final equation that includes two independent quantities $g_{n}$ and $h_{n}$ that are defined in terms of $ \mathcal V_{nk}$ matrix elements and should be zero if $\{\hat \gamma_5, H_0 \} = 0$ is satisfied. 
The difference of susceptibilities is written then in terms of the eigenvalues $\lambda_n$ as a sum of two blocks:
\begin{equation}
  \Delta = \frac{1}{V(1-m^2)^2} \sum_n
  \frac{2m^2(1-\lambda_n^2)^2}{\lambda_n^4}
  +\frac{1}{V(1-m)^2} \sum_n 
  \left[\frac{h_{n}}{\lambda_n}-\frac{4g_{n}}{\lambda_n}\right].
  \label{eq:DeltaViol}
\end{equation}

\begin{figure}[t]
\centering
  \includegraphics[width=.8\textwidth]{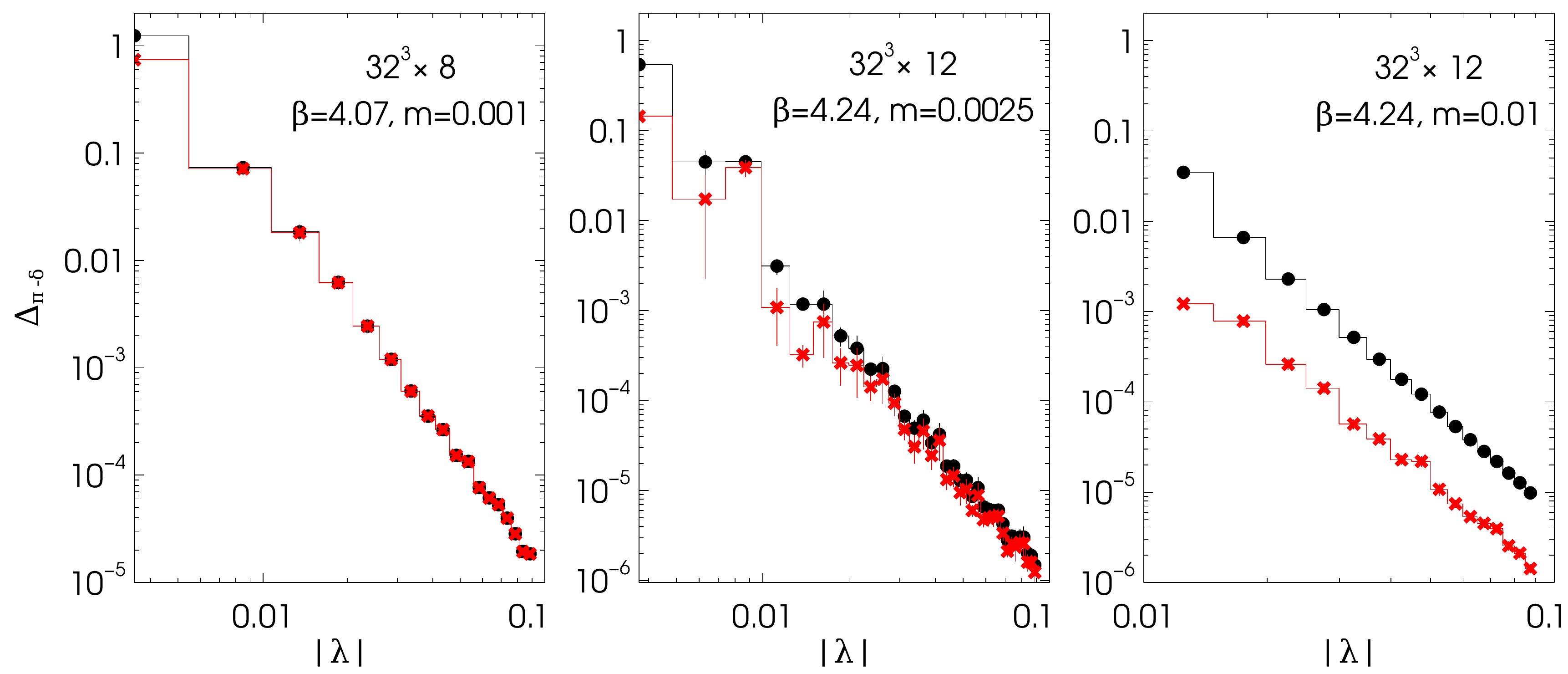}
  \caption{
    Eigenmode decomposition 
    of the suseptibility difference $\Delta$ for comparable temperatures $T\sim~1.1 T_c$.
    Average in each bin of $\lambda$ is plotted for full contribution 
    (black circles) and for the contribution from the GW violating
    term (red crosses).
  }
  \label{fig:DeltaHnn}
\end{figure}

This exact decomposition allows a quantitative measure of the contribution to $\Delta$ of the discretization artifacts that violate the Ginsparg-Wilson. These are the terms including $g_{n}$ and $h_{n}$. It turns out that for the coarsest lattices, $N_t=8$, the violation terms account for about 60\% of the total signal, up to 97\% in the worst cases. The finer lattices, $N_t=12$, show typical deviations of less than 10\% with peaks of 30\%. The susceptibility $\Delta$ is highly sensitive to chiral symmetry violation in the lowest part of the spectrum as seen in Figure~\ref{fig:DeltaHnn}. We show there the spectral decomposition of the terms in the sum eq.~(\ref{eq:DeltaViol}) separating the full sum $\Delta$ from the GW violating contribution, black circles and red crosses respectively. It is evident that the low modes are the main source of the signal and main source of the violation at the same time (notice the logarithmic scale).

\subsection{Reweighting}

Motivated by the previous results we reweight the results from our action to the one that satisfies the GW relation exactly. We call it the overlap action below, but it has the same kernel as the domain-wall fermion. The sign function approximation is improved by treating the low-lying eigenvalues of the kernel exactly. We calculate the reweighting factor as described in \cite{Fukaya:2013vka}. We can also perform the measurements on our M\"obius domain wall fermion configurations using overlap valence quarks. This procedure is called partial quenching since the Dirac operator in the partition function has different discretization than the one used for the measurement. 

\begin{figure}[b]
  \centering
  \includegraphics[width=.4\textwidth]{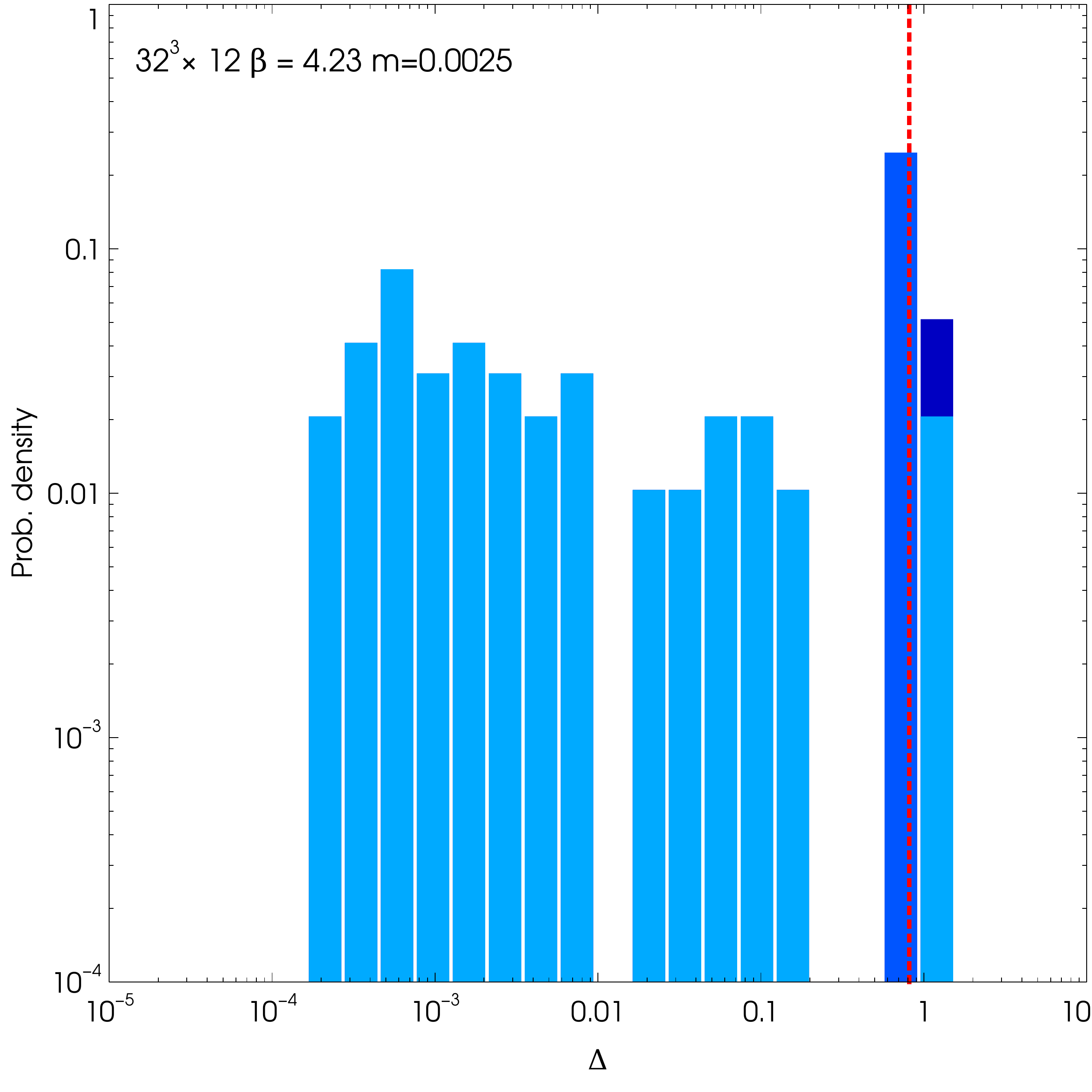}
  \includegraphics[width=.4\textwidth]{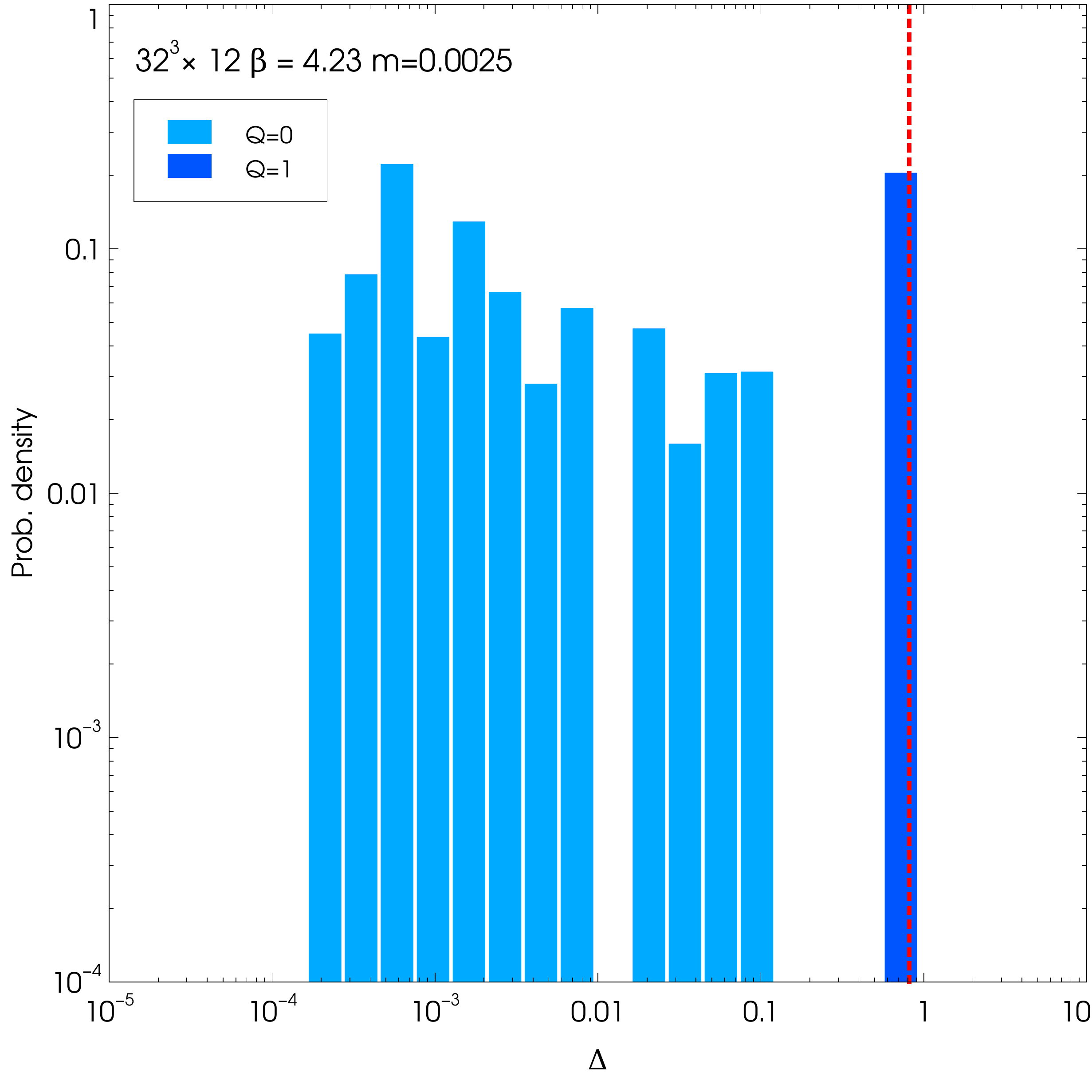}
\caption{Plot of the histograms showing the partially quenched result (left) and the reweighted one (right) for the $\Delta$ reconstructed from the overlap operator spectrum. The ensemble is $32^3\times 12$, $\beta=4.23$, $am=0.0025$, in physical units translates to $T/T_c \sim 1.05$ with an estimated pion mass of about 210 MeV.}
\label{fig:figure/RWcompare}
\end{figure}

The histogram of $\Delta$ from the overlap operator spectral sum is shown in Figure~\ref{fig:figure/RWcompare}, left. The colors differentiate the topological sectors. The lightest color indicates the zero topology sector that is expected to survive in the thermodinamical limit. The corresponding reweighted histogram of $\Delta$ is shown in the right panel. Notice that the rightmost peak in the Q=0 sector present in the partially quenched plot disappears after the procedure. This peak is coming from near zero modes that have poor chiral symmetry properties. These modes are mostly localized in few lattice spacings, another indication that they are lattice artifacts. 

This observation, common to other ensembles, suggests that partial quenching could potentially overestimate the result for $\Delta$ and thus reweighting is essential to get a result that respects chiral symmetry. 

\begin{figure}[ptb]
  \centering
  \includegraphics[width=.4\textwidth]{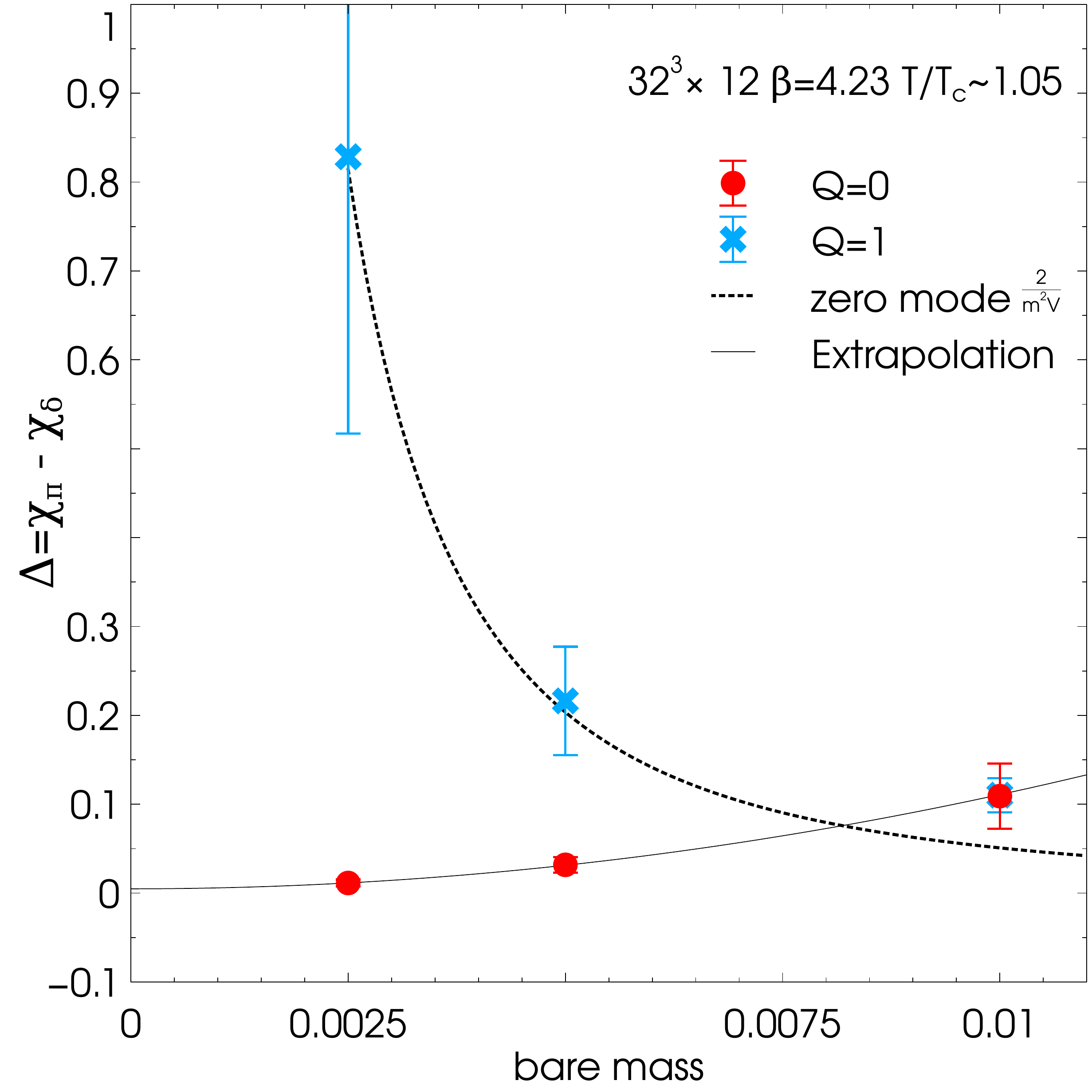}
  \includegraphics[width=.4\textwidth]{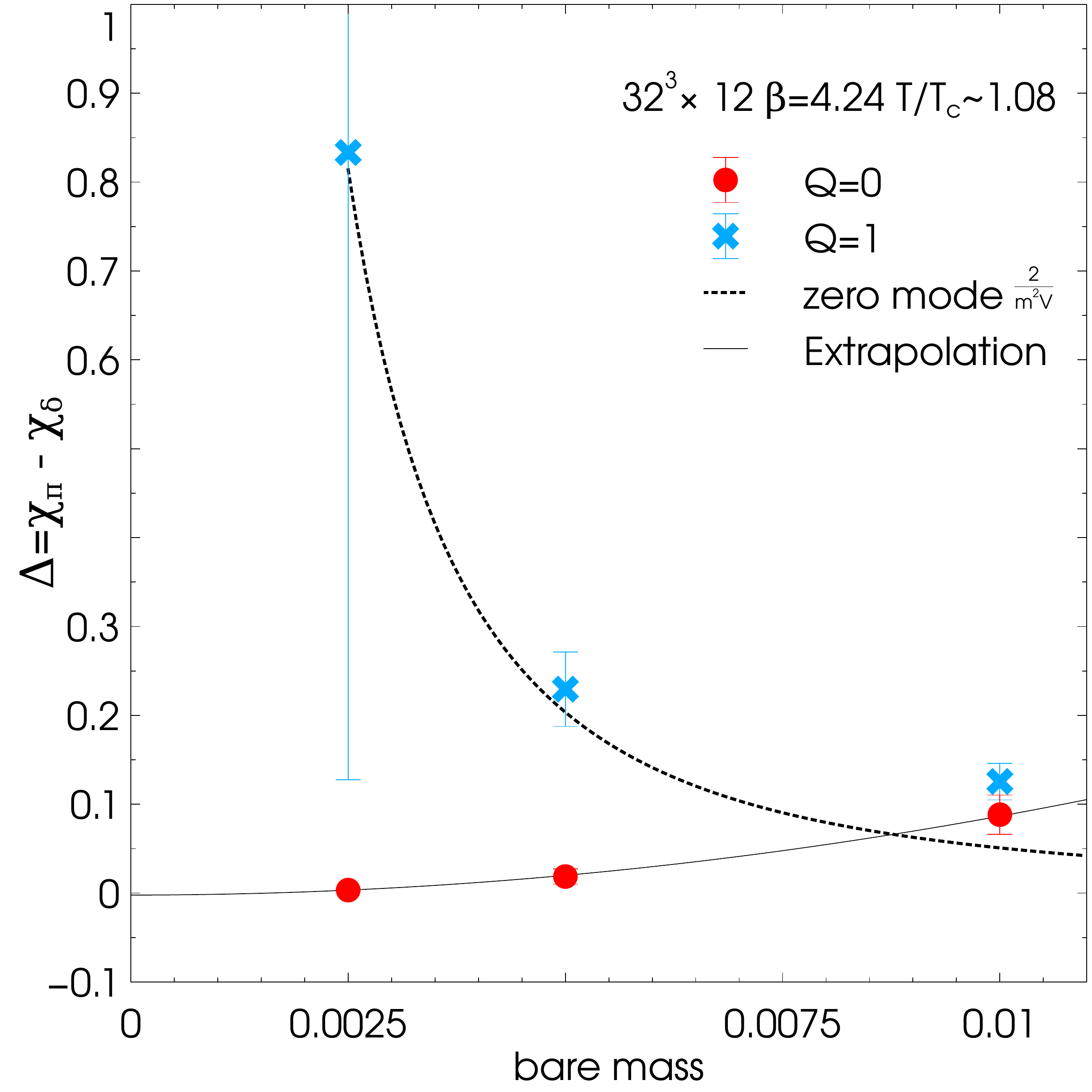}
\caption{Mass dependence of the reweighted $\Delta$ for two temperatures close to the phase transition in the finer lattices. The topological sector averages are separately plotted. The $Q=0$ averages are extrapolated to a result compatible to zero in the chiral limit.}
\label{fig:figure/FinalAfterRW}
\end{figure}

Collecting the reweighted averages we show the mass dependence for two temperatures above the phase transition in Figure~\ref{fig:figure/FinalAfterRW}. We separate the averages for different topological sectors and show only the one corresponding to $Q=0,1$.  In the Q=1 sector the central values are very close to the predicted $1/m^2$ dependence, indicating that the signal is essentially coming from the single zero mode. The bulk of the sum in eq.~(\ref{eq:DeltaSpectr}) is subdominant. As stated in the previous section, the zero-mode term in the non trivial sector is expected to have a vanishing contribution in the thermodinamical limit. We thus consider the $Q=0$ sector averages as and example of the physical contribution surviving in the large volume limit. The plots for two temperatures close to the phase transition show a suppression in the chiral limit of the difference of susceptibilities. The continuous lines represent an extrapolation assuming a simple quadratic form. The extrapolated result is compatible with zero within errors. It implies that the physical contribution to the axial symmetry breaking disappears in the thermodinamical and chiral limit just above the phase transition. 

\section{Conclusion}

The difference of susceptibilities $\Delta= \chi_\pi - \chi_\delta$ is a probe for the axial $U(1)$ symmetry. If $\Delta$ is zero then the axial symmetry is invisible in the scalar meson channels and it appears restored in that sector of the particle spectrum. We show that the measurement of $\Delta$ is delicate and highly sensitive to the level of chiral symmetry in the lattice action. A direct measure of the effect of the artifacts to the averages of $\Delta$ indicates that in the coarse lattices ($N_t=8$, about 0.11 fm near the phase transition) they account for the major part of the signal. The effect is reduced with the lattice spacing but still can generate a significant contribution to $\Delta$. We reweight the results to an action that satisfies exactly the Ginsparg-Wilson relation. We show that a partially quenched measure of $\Delta$ using the overlap operator can still be affected by artifacts in the low-mode region in comparison to the corresponding reweighted result. The $\Delta$ averages after the reweighting procedure show a mass dependence that is compatible with zero in the chiral limit. Our current conclusion is that the axial anomaly at $T \gtrsim T_c$ is invisible in the scalar meson channels. This result is compatible with the previous conclusion using overlap fermions \cite{Cossu:2013uua}. A detailed account of the measurement will be reported in a series of papers in preparation.  

\acknowledgments 
Numerical simulations are performed on IBM System Blue Gene Solution at the High Energy Accelerator Research Organization (KEK) under the Large Scale Simulation Program (No. 13/14-4, 14/15-10). We thank P. Boyle for helping in the optimization of the code for BGQ. This work is supported in part by the Grant-in-Aid of the Japanese Ministry of Education (No. 25800147, 26247043, 15K05065) and  SPIRE (Strategic Program for Innovative Research) Field 5.

\bibliographystyle{sissa}
\bibliography{proceedingsU1.bib}

\end{document}